\documentclass[aps,prl,twocolumn, draft, showpacs,intlimits,amsmath,amssymb,floats]{revtex4-1}
\usepackage{bm}
\usepackage[final]{graphicx}
\usepackage{epsfig}

\begin{document}

\title{Theory of two-magnon Raman scattering in iron pnictides and chalcogenides}

\author{C.-C. Chen$^{1,2}$}
\author{C. J. Jia$^{1,3}$}
\author{A. F. Kemper$^{1}$}
\author{R. R. P. Singh$^4$}
\author{T. P. Devereaux$^{1,5}$}

\address{$^1$Stanford Institute for Materials and Energy Science, SLAC National Accelerator Laboratory, 2575 Sand Hill Road, Menlo Park, California 94025, USA}
\address{$^2$Department of Physics, Stanford University, Stanford, California 94305, USA}
\address{$^3$Department of Applied Physics, Stanford University, Stanford, California 94305, USA}
\address{$^4$Department of Physics, University of California, Davis, California 95616, USA}
\address{$^5$Geballe Laboratory for Advanced Materials, Stanford University, California 94305, USA}

\date{\today}

\begin{abstract}
Although the parent iron-based pnictides and chalcogenides are itinerant antiferromagnets, the use of local moment picture to understand their magnetic properties is still widespread. We study magnetic Raman scattering from a local moment perspective for various quantum spin models proposed for this new class of superconductors. These models vary greatly in the level of magnetic frustration and show a vastly different two-magnon Raman response. Light scattering by two-magnon excitations thus provides a robust and independent measure of the underlying spin interactions. In accord with other recent experiments, our results indicate that the amount of magnetic frustration in these systems may be small.
\end{abstract}

\maketitle
Since their discovery in 2008~\cite{Hosono, MKWu}, iron-based superconductors have led to extensive studies in the condensed matter field. Like the high-$T_c$ cuprate superconductors, the iron pnictides and chalcogenides are quasi-two-dimensional materials with a layered structure. Both the copper and iron based superconductors exhibit antiferromagnetic (AF) order in the parent phases, and superconductivity emerges by suppressing magnetism upon doping. However, there are also clear differences. At stoichiometry, the cuprates are Mott insulators and can be described by an effective single-band model. On the other hand, the iron-based pnictide and chalcogenide parent compounds are multi-orbital systems and remain metallic even in the AF state. The description of magnetism and its interplay with superconductivity in this new class of superconductors remain an active area of research.

There is growing evidence that weak coupling theories with varying degrees of sophistication and input from density functional theory (DFT) calculations can quantitatively describe several experimental findings~\cite{Mazin, Kuroki, Kemper}. However, many aspects of the magnetic properties in these materials are also remarkably captured by a local moment perspective. These include the phase diagram with orthorhombic distortion and antiferromagnetic order, temperature dependence of uniform susceptibility, and neutron studies of spin-wave dispersion throughout the Brillouin zone (BZ). It also has been shown that a local moment model can reproduce the essence of magnetism from DFT calculations~\cite{Antropov}. There are various reasons why a strong coupling perspective is still relevant. It provides the potential for a unified framework for understanding high-$T_c$ superconductivity derived from electron-electron interactions. More importantly, \emph{Local moment models are being widely used in experiments} to explain magnetic properties in these systems~\cite{Dai_Ca122, Ba122_Ewings, Sr122_Ewings, Sugai_122_1, Sugai_122_2, Neutron_Dai_FeTe, Sugai_FeTe}.

In the pnictides, proposed theories for their collinear AF phase based on localized moments include the spatially anisotropic $J_{1a}$-$J_{1b}$-$J_2$ model~\cite{J1abJ2_1, J1abJ2_2}, where the coupling is ferromagnetic (FM) in one direction and AF in the other, and the strongly frustrated $J_1$-$J_2$ model~\cite{J1J2_1, J1J2_2, J1J2_3, J1J2_4}, where collinear antiferromagnetism arises via order by disorder (see Fig. 1). Similarly, the diagonal double stripe AF order in iron chalcogenides can be obtained by either invoking strong frustration in a $J_1$-$J_2$-$J_3$ model~\cite{Hu_J1J2J3} or utilizing a model with strong spatial anisotropy stemming from orbital order~\cite{Ashvin_SO}. These Hamiltonians can lead to ground states of the same broken symmetry, but they vary greatly in the degree of magnetic frustration. It is necessary to distinguish these scenarios and narrow the possible models for the iron-based superconductors.

The preceding difficulty in some cases could be resolved by inelastic neutron scattering (INS) experiments. Indeed, the INS spin wave spectrum of CaFe$_2$As$_2$~\cite{Dai_Ca122} has a local maximum at momentum $(\pi,\pi)$ (in the single iron BZ), strongly favoring the $J_{1a}$-$J_{1b}$-$J_2$ model. However, as high energy spin wave signals are strongly damped, interpretations of the neutron results remain disputed~\cite{Neutron_itinerant}. Moreover, there are also cases where different Hamiltonians can lead to virtually indistinguishable spin wave spectra. Therefore, further studies based on different measurements appear crucial.

One way to achieve the above goal is by studying magnetic Raman scattering~\cite{RMP}. Light scattering by two-magnon flips is dominated by short-range excitations and sensitive to details of the exchange couplings. This probe was instrumental in the first accurate determination of the exchange constants in cuprate superconductors~\cite{Raman_cuprates}.

\begin{figure}[t!]
\includegraphics[width=\columnwidth]{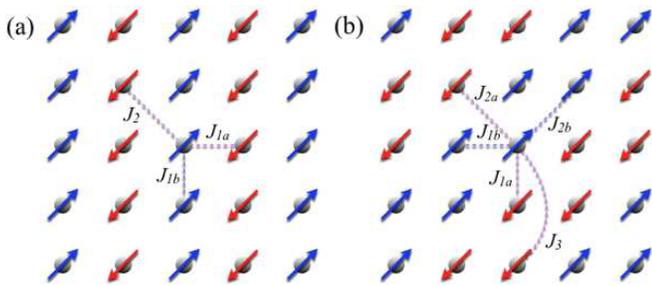}
\caption{
(Color online) Schematics of quantum spin models that lead to (a) the $(\pi,0)$ collinear AF order in iron pnictides, and (b) the $(\pi/2,\pi/2)$ diagonal double stripe AF order in iron chalcogenides.
}\label{fig:Cartoon}
\end{figure}

In the present study, following the Fleury-Loudon (FL) formalism~\cite{FL_formalism,SS_formalism}, we investigate the two-magnon Raman spectra for strongly frustrated models and spatially anisotropic models that are unfrustrated. Using exact diagonalization (ED) and spin wave theory, we find that in strongly frustrated models the two-magnon peak appears at relatively low energies. Strong frustration implies abundant low-energy configurations, which thereby provides symmetry-dependent pathways for low-energy two-magnon flips. In contrast, in unfrustrated models the two-magnon peak occurs at a higher energy, closer to twice the single-magnon bandwidth. Magnetic Raman scattering thus can be a robust and independent measure of the underlying spin interactions.

We study quantum spin models on a square lattice:
\begin{equation}\label{eq:Model}
H=\sum_{ij}\frac{J_{ij}}{2}\mathbf{S}_i\cdot\mathbf{S}_{j},
\end{equation}
where $J_{ij}$ are short range exchange couplings up to the third nearest-neighbor (NN). Depending on the interaction parameters, the model can support ground states of different broken symmetries, such as the $(\pi,0)$ collinear AF order or the $(\pi/2,\pi/2)$ diagonal double stripe order. The FL Raman scattering operator $\hat{\cal{O}}$ is given by~\cite{RMP}
\begin{equation}\label{eq:FL_operator}
\hat{\cal{O}}=\sum_{i,j} \eta_{ij} (\hat{\mathbf{e}}_{in}\cdot  \hat{\mathbf{d}}_{ij})(\hat{\mathbf{e}}_{out}\cdot \hat{\mathbf{d}}_{ij})\mathbf{S}_i\cdot\mathbf{S}_j,
\end{equation}
where the coupling strengths $\eta_{ij}$ are proportional to the exchange interactions $J_{ij}$, $\hat{\mathbf{d}}_{ij}$ is the unit vector that connects lattice sites $i$ and $j$, and $\hat{\mathbf{e}}_{in}$ and $\hat{\mathbf{e}}_{out}$ are the incident and scattered photon polarizations, respectively. Specifically, we will consider the following light-polarization geometries:
\begin{eqnarray}
B_{1g}:&&\hat{\mathbf{e}}_{in}=\frac{1}{\sqrt{2}}(\hat{x}+\hat{y}),\hat{\mathbf{e}}_{out}=\frac{1}{\sqrt{2}}(\hat{x}-\hat{y}),
\nonumber\\
A_{1g}^{\prime}:&&\hat{\mathbf{e}}_{in}=\frac{1}{\sqrt{2}}(\hat{x}+\hat{y}),\hat{\mathbf{e}}_{out}=\frac{1}{\sqrt{2}}(\hat{x}+\hat{y}),
\end{eqnarray}
where we note that $A^{\prime}_{1g}$ transforms as $A_{1g}\oplus B_{2g}$~\cite{RMP}. With a given scattering operator $\hat{\cal{O}}$, the two-magnon Raman cross-section is given by $R(\omega)=-(1/\pi)\textrm{Im}[I(\omega)]$, where
\begin{equation}\label{eq:Raman}
I(\omega)\equiv \langle \Psi_0 \vert \hat{\cal{O}}^\dagger \frac{1}{\omega+E_0+i\epsilon -H}  \hat{\cal{O}}\vert\Psi_0\rangle.
\end{equation}
Here $\vert \Psi_0\rangle$ is the ground state with energy $E_0$.


Below we perform spin one-half ($S=1/2$) calculations in systems with a $(\pi,0)$ collinear AF order, and spin one ($S=1$) calculations in systems with a $(\pi/2,\pi/2)$ diagonal double stripe order. These calculations are relevant since in the pnictides the Fe moment is unexpectedly small (ranging from 0.3 to 0.9 $\mu_B$), while in the chalcogenides the moments ($\sim 2\mu_B$) are much larger~\cite{David_Review}. For the $S=1/2$ case, we perform ED calculations on square lattices of 16, 32, and 36 sites. For the $S=1$ case, we focus on a 16-site cluster as the Hilbert space is substantially larger. We further compare the calculations with spin wave theory and comment on the effects of $S$.

We first discuss the case of ($\pi,0$) AF order. We consider (i) a frustrated $J_1-J_2$ model with $J_1=J_2$ and (ii) the spatially anisotropic $J_{1a}$-$J_{1b}$-$J_2$ model, as depicted in Fig. \ref{fig:Cartoon}(a). For the latter we take $J_{1b}=-0.1J_{1a}$ and $J_2=0.4J_{1a}$ as reported by INS measurements on CaFe$_2$As$_2$~\cite{Dai_Ca122}. In this case, even though $(J_{1a}+J_{1b})$ is comparable to $2J_2$, the system remains unfrustrated.

\begin{figure}[t!]
\includegraphics[width=\columnwidth]{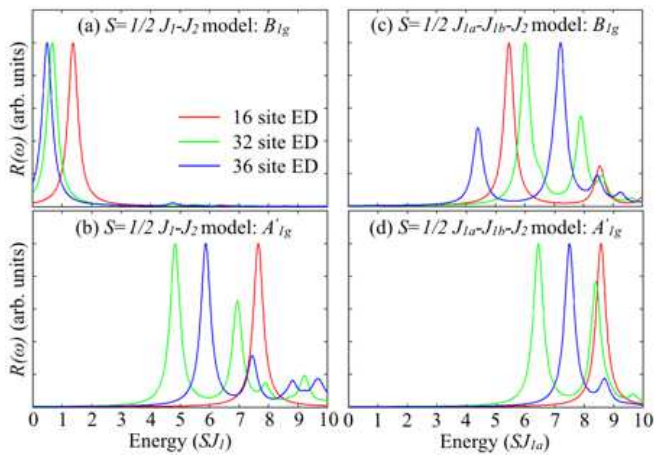}
\caption{
(Color online) The Fleury-Loudon two-magnon Raman cross-sections calculated with $S=1/2$ exact diagonalization. In the frustrated $J_1$-$J_2$ model, the $B_{1g}$ two-magnon excitation appears at $\sim SJ_1$, an energy much lower than twice the single-magnon bandwidth.
}\label{fig:Raman_CAF}
\end{figure}

Figure \ref{fig:Raman_CAF} shows the two-magnon Raman spectra. In the 36-site ED calculations, the $J_1$-$J_2$ model has a $B_{1g}$ peak at $\sim SJ_1$, while in the $A_{1g}^{\prime}$ channel it is at $\sim 6SJ_1$. On the other hand, for the $J_{1a}$-$J_{1b}$-$J_2$ model the two-magnon excitation appears at $\sim 7SJ_{1a}$ in both polarizations. We note that while finite-size effects give small corrections to the Raman resonance energy, the clear qualitative difference in the $B_{1g}$ channel between the two models is robust, depending weakly on cluster sizes.

The sharp distinction in the $B_{1g}$ two-magnon response is related to the difference in the amount of magnetic frustration. In both models, the single-magnon bandwidth is roughly the same~\cite{spin-orbital}; however, their magnon spectral weights are rather distinct. In the $J_{1a}$-$J_{1b}$-$J_2$ model, the magnon density of states (DOS) diverges at the top of the single-magnon band, similar to that in the Heisenberg model. On the other hand, in the frustrated $J_1$-$J_2$ model a large spectral weight is distributed at lower energies~\cite{impurity}. In general, a substantial low-energy magnon DOS as implied by frustration provides pathways for low-energy two-magnon flips~\cite{triangular_1, triangular_2}.

The low-energy $B_{1g}$ mode of the $J_1$-$J_2$ model can be related to the proximity of the system to a disordered phase characterized by either a quantum spin liquid or a valence bond solid (VBS). At a transition to a phase lacking magnetic order, one expects singlet modes with appropriate quantum numbers to soften to zero energy~\cite{Lhuillier}. Close to such a critical point low energy excitations would appear. 
For a quantum spin liquid, low energy singlets may appear in all light polarizations, while for a VBS phase low energy singlets may appear only in specific channels. Our finding of a low-energy excitation in $B_{1g}$ and not in $A^{\prime}_{1g}$ implies a nearby VBS phase with columnar dimers, since this state has the right symmetry~\cite{Dimer,ED_36}.

To compare the calculations to experiments, we convert our results into physically measurable units. By fitting to the $J_{1a}$-$J_{1b}$-$J_2$ model, an $SJ_{1a}\sim50$ meV has been reported by neutron scattering experiments on CaFe$_2$As$_2$~\cite{Dai_Ca122}. Therefore, the 36-site ED calculation indicates a $B_{1g}$ two-magnon peak at $7.4J_{1a}\sim3000$ cm$^{-1}$. Anisotropic exchange couplings are also found in BaFe$_2$As$_2$~\cite{Ba122_Ewings} and SrFe$_2$As$_2$~\cite{Sr122_Ewings}, with a slightly different energy scale. Even with a material dependence and possible corrections due to finite-size calculations, the two-magnon peak in an unfrustrated $J_{1a}$-$J_{1b}$-$J_2$ model is clearly at an energy several times higher than $SJ_{1a}$. On the other hand, for a $J_1$-$J_2$ model with the same exchange energy scale, a $B_{1g}$ two-magnon peak occurs around a few hundred wavenumbers. As in the cuprates~\cite{J1J2Raman_1990}, the absence of a low-energy $B_{1g}$ two-magnon peak in recent experiments implies that magnetic frustration in the pnictides might also be small~\cite{Sugai_122_1, Sugai_122_2}.


To further understand the Raman cross-sections, we discuss results based on spin wave theory. In the AF Heisenberg model, linear spin-wave theory indicates a $B_{1g}$ two-magnon peak located at $8SJ$, twice the energy of the single-magnon bandwidth. This value is appropriate only in the classical $S=\infty$ limit, where magnon-magnon interactions can be neglected. In systems with a finite $S$, two magnons are bound to each other locally in space, thereby shifting the two-magnon energy to a lower value. When interaction effects are treated at the mean-field level, the two-magnon Raman response takes the following RPA form: $R(\omega) \sim \textrm{Im} \left[I(\omega)/(1+\frac{\bar{J}}{S}I(\omega))\right]$, 
where $I(\omega)$ [defined in Eq. (\ref{eq:Raman})] is calculated within the linear spin wave approximation, and $\bar{J}$ is the mean-field interaction characteristic to the problem. In the $S=1/2$ AF Heisenberg model, the $B_{1g}$ two-magnon energy is renormalized to $\sim 6SJ$. Similarly, in the $S=1/2$ ED calculations for the $J_{1a}$-$J_{1b}$-$J_2$ model, the two-magnon excitation in the $B_{1g}$ channel occurs at $\sim 7SJ_{1a}$, again understood by an RPA renormalization of its classical $B_{1g}$ two-magnon peak at $8SJ_{1a}$. 

In frustrated models, however, the correspondence between the classical $B_{1g}$ two-magnon energy and twice the single-magnon bandwidth fails to describe the Raman spectra. Based on linear spin wave theory, while the top of the single magnon band is roughly $6SJ_1$ in the $J_1$-$J_2$ model, its classical $B_{1g}$ two-magnon energy occurs at zero energy. This zero energy two-magnon resonance is closely related to the zero mode in the spin wave spectrum at momentum $(\pi,\pi)$, which is known to be shifted to a finite energy by quantum fluctuations. 
The spin-wave theory calculation agrees well with the ED results shown in Fig. \ref{fig:Raman_CAF}, where the $B_{1g}$ two-magnon energy $\sim SJ_1$ is much closer to zero, rather than twice the single-magnon bandwidth. Moreover, our $S=1$ 16-site ED calculation indicates that the $B_{1g}$ two-magnon peak moves further down to $\sim0.7SJ_1$. The distinction in the two-magnon response thus serves as a clear benchmark to distinguish the $J_1$-$J_2$ and the $J_{1a}$-$J_{1b}$-$J_2$ models. 

We next focus on situations where the model Hamiltonian [Eq. (\ref{eq:Model})] has a $(\pi/2,\pi/2)$ AF ground state as observed in iron chalcogenides~\cite{p2p2_1,p2p2_2}. Like the case of the pnictides, below we discuss two models that have a similar single-magnon bandwidth but vary in the amount of magnetic frustration.

In the first case we consider a system which is spatially anisotropic but unfrustrated. We use parameters from DFT calculations~\cite{J1abJ2_2} (in units of meV): $SJ_{1a}=-7.6$, $SJ_{1b}=-26.5$, $SJ_{2a}=46.5$, and $SJ_{2b}=-34.9$ [see Fig. \ref{fig:Cartoon}(b)]. In the second frustrated case, we use parameters from  Fe$_{1.05}$Te INS measurements~\cite{Neutron_Dai_FeTe} (in units of meV): $SJ_{1a}=-17.5$, $SJ_{1b}=-51.0$, $SJ_{2a}=21.7$, $SJ_{2b}=21.7$, and $SJ_3=6.8$. In the latter case a strong frustration is present, as the NN interactions in both crystal-axes are FM, and the second NN exchanges are all AF, independent of spin sub-lattice. In both cases, $SJ$ below is referred to as the energy scale of the dominant coupling.

\begin{figure}[t!]
\includegraphics[width=\columnwidth]{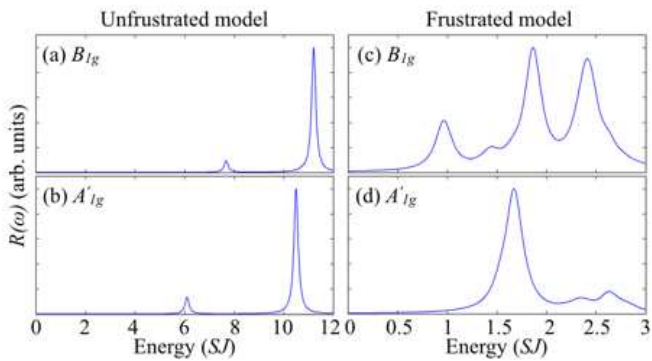}
\caption{
(Color online) Two-magnon Raman spectra for models with a $(\pi/2,\pi/2)$ diagonal double stripe AF order. In both cases $SJ$ is the energy of the dominant exchange coupling. We note the plots have a  different energy scale.
}\label{fig:Raman_FeTe}
\end{figure}

\begin{figure}[t!]
\includegraphics[width=\columnwidth]{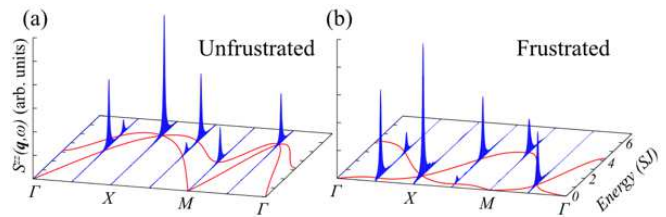}
\caption{
(Color online) $S^{zz}(\mathbf{q},\omega)$  [blue peaks] from ED and the magnon dispersion spectra [red curves] from spin wave theory. Here the momentum points are (in the single iron BZ): $\Gamma =(0,0)$, $X=(\pi,0)$, and $M=(\pi,\pi)$.
}\label{fig:Spinwave_FeTe}
\end{figure}

Figure \ref{fig:Raman_FeTe} shows the two-magnon Raman spectra from 16-site $S=1$ ED calculations. In the unfrustrated system, there is no low-energy two-magnon excitation. The two-magnon Raman spectra start to show spectral features between $6\sim 8 SJ$, and the cross-sections are dominated by peaks located at $\sim 11 SJ$ and $10.5 SJ $ respectively in the $B_{1g}$ and $A^{\prime}_{1g}$ polarizations  [Fig. \ref{fig:Raman_FeTe}(a) and (b)]. 
In this unfrustrated case, the two-magnon peak is expected to occur at an energy close to twice the single-magnon bandwidth [see Fig. \ref{fig:Spinwave_FeTe}]. On the other hand, when the system is frustrated there is a substantial low-energy two-magnon weight between $1\sim3 SJ$ [Fig. \ref{fig:Raman_FeTe}(c) and (d)] 
These results show a clear difference in the two-magnon Raman response between the two models. 

As mentioned previously, the distinction in the two-magnon Raman spectra is not caused by a difference in the single-magnon bandwidth. As shown in Fig. \ref{fig:Spinwave_FeTe}, the magnon dispersion from spin wave theory~\cite{Hu_J1J2J3, Neutron_Dai_FeTe} and the dynamical structure factor $S^{zz}(\mathbf{q},\omega)$ from ED calculations both indicate a similar magnon bandwidth of the two models. However, in a frustrated model the magnon DOS is no longer dominated by zone boundary magnons, and an estimate based on twice the magnon bandwidth fails to describe the two-magnon Raman profile.

We last note that recent Raman scattering experiments do not suggest the existence of low-energy two-magnon excitations in the chalcogenides~\cite{Sugai_FeTe}. This implies that weakly frustrated but spatially anisotropic models based on orbital order~\cite{Ashvin_SO} might be relevant. When there is an ambiguity or freedom in fitting the neutron scattering spin wave spectra to localized model models, additional experiments such as Raman scattering can be helpful for pinning down the interaction parameters. 

In summary, we have calculated the two-magnon Raman spectra for various spin models proposed for the iron-based superconductors. We have shown that a distinct two-magnon Raman response can result between models that vary in the level of magnetic frustration. Complementary to neutron scattering, magnetic Raman scattering detects short-wavelength spin fluctuations and can serve as an independent measure of the underlying spin interactions. Together with recent experiments, our results favor spatially anisotropic models, implying that the amount of magnetic frustration in the pnictides and chalcogenides is small.


The authors acknowledge discussions with M. Bernhard, R. Hackl, M. Gingras, J.-H. Chu, and P. Hubbard. CCC, CJJ, AFK, and TPD are supported by the U.S. DOE under Contract No. DE-AC02-76SF00515.  CJJ is also supported by the Stanford Graduate Fellowships Program. RRPS is supported by NSF Grant No. DMR-1004231. This research used resources of NERSC, supported by U.S. DOE under Contract No. DE-AC02-05CH11231.


\begin{thebibliography}{arprev}

\bibitem{Hosono}
Y. Kamihara \emph{et al}., J. Am. Chem. Soc. \textbf{130}, 3296 (2008).

\bibitem{MKWu}
F. C. Hsu \emph{et al.}, Proc. Natl. Acad. Sci. \textbf{105}, 14262 (2008).

\bibitem{Mazin}
I. I. Mazin \emph{et al.}, Phys. Rev. Lett. \textbf{101}, 057003 (2008).

\bibitem{Kuroki}
K. Kuroki \emph{et al.}, Phys. Rev. B \textbf{79}, 224511 (2009).

\bibitem{Kemper}
 A. F. Kemper \emph{et al.}, New J. Phys. \textbf{12} 073030 (2010).

\bibitem{Antropov}
A. L. Wysocki \emph{et al.}, arXiv:1011.1715.

\bibitem{Dai_Ca122}
J. Zhao {\it et al.}, Nat. Phys. \textbf{5}, 555 (2009).

\bibitem{Ba122_Ewings}
R. A. Ewings {\it et al.}, Phys. Rev. B \textbf{78}, 220501(R) (2008).

\bibitem{Sr122_Ewings}
R. A. Ewings {\it et al.}, arXiv:1011.3831.

\bibitem{Sugai_122_1}
S. Sugai {\it et al.}, Phys. Rev. B \textbf{82}, 140504(R) (2010).

\bibitem{Sugai_122_2}
S. Sugai {\it et al}., arXiv:1010.6151.

\bibitem{Neutron_Dai_FeTe}
O. J. Lipscombe {\it et al.}, arXiv:1011.2930.

\bibitem{Sugai_FeTe}
K. Okazaki {\it et al.}, Phys. Rev. B \textbf{83}, 035103 (2011).

\bibitem{J1abJ2_1}
Z. P. Yin \emph{et al.}, Phys. Rev. Lett. \textbf{101}, 047001 (2008).

\bibitem{J1abJ2_2}
M. J. Han \emph{et al}., Phys. Rev. Lett. \textbf{102}, 107003 (2009).

\bibitem{J1J2_1}
T. Yildirim, Phys. Rev. Lett. \textbf{101}, 057010 (2008).

\bibitem{J1J2_2}
Q. Si and E. Abrahams, Phys. Rev. Lett. \textbf{101}, 076401 (2008).

\bibitem{J1J2_3}
C. Fang \emph{et al.}, Phys. Rev. B \textbf{77}, 224509 (2008).

\bibitem{J1J2_4}
C. Xu, M. M\"uller, and S. Sachdev, Phys. Rev. B \textbf{78}, 020501(R) (2008).

\bibitem{Hu_J1J2J3}
C. Fang, B. A. Bernevig, and J. P. Hu, Euro. Phys. Lett. \textbf{86}, 67005 (2009).

\bibitem{Ashvin_SO}
A. M. Turner, F. Wang, and A. Vishwanath, Phys. Rev. B \textbf{80}, 224504 (2009).

\bibitem{Neutron_itinerant}
S. O. Diallo {\it et al.}, Phys. Rev. Lett. \textbf{102}, 187206 (2009).

\bibitem{RMP}
T. P. Devereaux and R. Hackl, Rev. Mod. Phys. \textbf{79}, 175 (2007).

\bibitem{Raman_cuprates}
R. R. P. Singh \emph{et al.}, Phys. Rev. Lett. \textbf{62}, 2736 (1989).

\bibitem{FL_formalism}
P. A. Fleury and R. Loudon, Phys. Rev. \textbf{166}, 514 (1968).

\bibitem{SS_formalism}
B. S. Shastry and B. I. Shraiman, Phys. Rev. Lett. \textbf{65}, 1068 (1990).

\bibitem{David_Review}
David C. Johnston, Advances in Physics \textbf{59}, 803 (2010).

\bibitem{impurity}
C.-C. Chen {\it et al.}, arXiv:1010.2917.

\bibitem{spin-orbital}
C.-C. Chen \emph{et al.}, Phys. Rev. B \textbf{80}, 180418(R) (2009).

\bibitem{triangular_1}
F. Vernay, T. P. Devereaux, and M. J. P. Gingras, J. Phys.: Condens. Matter \textbf{19}, 145243 (2007).

\bibitem{triangular_2}
N. Perkins and W. Brenig, Phys. Rev. B \textbf{77}, 174412 (2008).

\bibitem{Lhuillier}
O. C\`{e}pas, J. O. Haerter, and C. Lhuillier, Phys. Rev. B \textbf{77}, 172406 (2008).

\bibitem{Dimer}
P. W. Leung, K. C. Chiu, and K. J. Runge, Phys. Rev. B \textbf{54}, 12938 (1996).

\bibitem{ED_36}
H. J. Schulz, T. A. L. Ziman, and D. Poilblanc, J. Phys. I \textbf{6}, 675 (1996).

\bibitem{J1J2Raman_1990}
F. Nori, E. Gagliano, and S. Bacci, Phys. Rev. Lett. \textbf{68}, 240 (1992).

\bibitem{p2p2_1}
W. Bao {\it et al.}, Phys. Rev. Lett. \textbf{102}, 247001 (2009).

\bibitem{p2p2_2}
S. Li {\it et al.}, Phys. Rev. B \textbf{79}, 054503 (2009). 

\end{thebibliography}
\end{document}